\documentclass[11pt, preprint]{article}
\usepackage{color} \usepackage{cancel}
\usepackage{float} \usepackage{graphicx}
\usepackage{subfigure} \usepackage{caption}	
\usepackage{dcolumn}
\usepackage[toc,page]{appendix} \usepackage{authblk}	 		
\usepackage{indentfirst}	
\usepackage{authblk}	 		
\usepackage{multirow}     
\usepackage{hyperref}     
\usepackage{amsfonts, amssymb, amsmath}
\usepackage[left=2.5cm,right=2.5cm,top=2.9cm,bottom=2.9cm,a4paper]{geometry}
\usepackage[noadjust]{cite}
\usepackage{filecontents}
\usepackage{soul}
\usepackage{xcolor}
\setstcolor{red}

\graphicspath{{./figures/}}
\hypersetup{ colorlinks   = true, 
urlcolor     = blue, 
linkcolor    = blue, 
citecolor    = red 	 
}

\title{\Large\bf Non-minimal coupling influence  on the deviation from de Sitter cosmological expansion}
\author[1]{I. V. Fomin\thanks{ingvor@inbox.ru}}
\author[2,3]{S. V. Chervon\thanks{chervon.sergey@gmail.com}}
\affil[1]{\small \it  Bauman Moscow State Technical University, 2-nd Baumanskaya street, 5, Moscow, 105005, Russia}
\affil[2]{\small \it Ulyanovsk State Pedagogical University, Ulyanovsk, Lenin's Square, B. 4/5, 432071, Russia}
\affil[3]{\small \it Kazan Federal University, Kremlevskaya street, 18, Kazan, 420008, Russia}

\begin{document}

\maketitle
\begin{abstract}
We investigate the models of cosmological inflation in generalized scalar-tensor gravity, which we consider as a source of deviation from de Sitter dynamics in the case of GR. Within the framework of the proposed approach, the exact equations of cosmological dynamics and parameters of cosmological perturbations are obtained.
\end{abstract}

\section{Introduction}
The inflationary paradigm is currently the main approach to describing the evolution of the early universe avoiding the horizon, flatness and homogeneity problems \cite{Starobinsky:1980te,Guth:1980zm,Linde:1981mu,Albrecht:1982wi}. Also inflation describes the formation of large scale structure of the universe in the context of the origin of primary inhomogeneities and relict gravitational waves \cite{Starobinsky:1979ty,Mukhanov:1981xt} (for review, see \cite{Mukhanov:1990me,Liddle}).
Most of the first inflationary models were based on the Einstein gravity coupled to self-interacting scalar field in Friedmann universe~\cite{Guth:1980zm,Linde:1981mu,Albrecht:1982wi} while the pioneering work by Starobinsky \cite{Starobinsky:1980te} was related to modified f(R) gravity which arose from quantum one-loop contributions of conformally covariant matter fields.

At the present time, to explain the stage of the accelerated expansion of the universe~\cite{Perlmutter:1998np,Riess:1998cb} (the construction of dark energy models), the cosmological models of modified f(R) gravity, which differ from Einstein gravity are actively considered
\cite{Elizalde:2004mq,Hu:2007nk,Appleby:2007vb,Starobinsky:2007hu,Nojiri:2008nt,Capozziello:2013xn}
(for review, see \cite{Nojiri:2017ncd}).
One of the oldest generalizations of GR, the scalar-tensor gravity (STG), also can describe  the early and later inflation \cite{Clifton:2011jh}.

Let us note that both f(R) gravity and STG are conformally related to GR with a scalar field minimally coupled to gravity.  The original papers where it was shown are \cite{Barrow:1988plb} for an arbitrary f(R) and \cite{Maeda:1989prd} for a more general scalar-tensor case.
Such conformal connection of the models with f(R) gravity and GR is considered, for example, in the work \cite{Motohashi:2017epjc}.
The conformal connection of cosmological models on the basis of STG and GR
have been studied for example, in \cite{DeFeliche:2011jcap}.
Therefore, on the basis of the inverse conformal transformation from GR to STG it is possible to find the connection between STG and f(R) gravity.
Such conformal connection between f(R) gravity and STG have been considered, for example, in
\cite{Sotiriou:2006cqg} and \cite{Faulkner:2006ub}.

Scalar-tensor gravity theories are important extensions of GR, which can explain both the initial inflationary evolution, as well as the late accelerating expansion of the Universe. The examples of inflationary models on the basis of scalar-tensor gravity theories with the exact solutions were considered in the works \cite{Demianski:2007mz, Demianski:2006pv, Belinchon:2016lwr, Belinchon:2017wvn, Belinchon:2015zrk,Faraoni,Fujii:2003pa}. Also, in the articles~\cite{Fomin_STFI,Fomin:2018blx} the equations of cosmological dynamics for STG were reduced to ones for the Einstein gravity in the case of the Friedman-Robertson-Walker metric by a specific choice of the coupling function and the kinetic function. This makes it easy to translate the solutions obtained for standard GR cosmology to the case of inflation based on STG. For this purpose it is possible to use the examples of exact solutions for inflation based on the Einstein gravity represented in~\cite{Chervon:2017kgn,Fomin:2017xlx}. Also it is of interest to pay attention to a new class of exact inflationary solutions in GR dubbed the constant-roll ones which found in
\cite{Motohashi:2015jcap}
and they compared with observations in
\cite{Motohashi:2017epl}.

In this paper, we consider the generalized scalar-tensor gravity theory and a deviation during inflationary stage from de Sitter expansion for such theories. This approach allow us to comply with the inflationary paradigm which implies quasi de Sitter expansion at an early stage of the evolution of the universe. Also we obtain the exact solutions of the cosmological dynamic equations and the parameters of cosmological perturbations in the generalized scalar-tensor gravity.

This paper is organized as follows. In section 2 we represent the generalized scalar-tensor (GST) theory which includes Brance-Dicke and induced gravity. We discuss the relation between scalar-tensor gravity and Einstein gravity with self-interacting scalar field. Also we write down cosmological equations for generalized scalar-tensor theory and note the transition to Friedmann cosmology. Section 3 is devoted to conformal connection GST cosmology with GR cosmology. In section 4 we introduce a connection (ansatz) of the Hubble parameter to non-minimal coupling function which fixes the deviation of Hubble parameter in GST cosmology from de Sitter stage in GR cosmology. In this way the non-minimal coupling function defines the evolution of a scalar field and the deviation of the potential from it in de Sitter stage. In section 5 we consider, as an example, the inflation with power-law Hubble parameter  and coupling function. The set of special class of exact solutions are obtained. The features of the slow-roll regime are considered. Section 6 is devoted to cosmological perturbations for the GST gravity under consideration. General relations for parameters of cosmological perturbations in terms of suggested ansatz are derived, including power spectrums, spectral indexes  and tensor-to-scalar ratio. Using observation data from PLANCK observatory we estimate the value of model parameter which gives a good correspondence to observations. In section 7 we discuss obtained results and conclude about advantages of STG gravity.

\section{Scalar Einstein-Friedmann and generalized scalar-tensor cosmologies}

We study the generalized scalar-tensor (GST) theory described by the action (see, for example,\cite{Faraoni}, Eq. (2.100))
\begin{eqnarray}\label{2.1}
S_{(GST)} = \frac{1}{\kappa}\int d^4x\sqrt{-g}\Big[\frac{1}{2}F(\phi) R -
 \frac{f(\phi)}{2}g^{\mu\nu}\partial_{\mu}\phi \partial_{\nu} \phi
 - V(\phi)\Big] +S^{(m)},\\
S^{(m)}=\int d^4x\sqrt{-g} L^{(m)},
\end{eqnarray}
where $ \kappa$ is the Einstein gravitational constant, $g$ a determinant of the spacetime metric $g_{\mu\nu}$, $\phi$ the scalar field with the potential $V=V(\phi)$, $f(\phi)$ and $F(\phi)$ are the differentiable functions of $\phi$, $R$ is the Ricci scalar curvature of the spacetime, $L^{(m)}$ is the Lagrangian of the matter. The action (\ref{2.1}) is the generalization associated with the scalar-tensor Brans-Dicke gravity where the linear interaction of  the scalar field $\phi $ to gravity is replaced by an coupling function $F(\phi) $ of the gravitational (non-material) scalar field. In the present article we will study the case of vacuum solutions for the model (\ref{2.1}) suggesting that $ S^{(m)}=0$.

Now, we note the actions for various partial cases of generalized scalar-tensor gravity theory:

1) Brans-Dicke gravity
\begin{eqnarray}\label{BD}
S_{(BD)} = \frac{1}{\kappa}\int d^4x\sqrt{-g}\Big[\frac{1}{2}\phi R -
 \frac{\alpha}{\phi}g^{\mu\nu}\partial_{\mu}\phi \partial_{\nu} \phi- V(\phi)\Big],\\
 F(\phi)=\phi,~~f(\phi)=\alpha/\phi,~~\alpha=const;
\end{eqnarray}
2) Induced gravity
\begin{eqnarray}\label{IND}
S_{(IND)} = \frac{1}{\kappa}\int d^4x\sqrt{-g}\Big[\frac{\beta}{2}\phi^{2} R -
\frac{f(\phi)}{2}g^{\mu\nu}\partial_{\mu}\phi \partial_{\nu} \phi- V(\phi)\Big],\\
 F(\phi)=\beta\phi^{2},~~\beta=const;
\end{eqnarray}
3) Non-minimal coupling
\begin{eqnarray}\label{NC}
S_{(NC)} = \frac{1}{\kappa}\int d^4x\sqrt{-g}\Big[(1-\xi\phi^{2}) R -
\frac{f(\phi)}{2}g^{\mu\nu}\partial_{\mu}\phi \partial_{\nu} \phi- V(\phi)\Big],\\
 F(\phi)=1-\xi\phi^{2}.
\end{eqnarray}
Here $\xi$ is the dimensionless coupling constant. Special values of this constant correspond to different cases:
$\xi=1/6$ (conformal coupling), $\xi=0$ (minimal coupling) and $\mid\xi\mid\gg 1$ (strong coupling).

The action of the scalar field Einstein gravity is
\begin{equation}\label{2.3}
  S_{(SFE)} = \int d^4x\sqrt{-g}\left[\frac{R}{2\kappa} -
 \frac{1}{2}g^{\mu\nu}\partial_{\mu}\varphi \partial_{\nu} \varphi
 - U(\varphi)\right].
\end{equation}
Let us remind that opposite to the scalar field $\phi$ in the action (\ref{2.1}), the scalar field $\varphi$ in the action (\ref{2.3}) is the source of the gravitation and it is a matter (material) field.

Considering the gravitational and field equations of the models (\ref{2.1}) and (\ref{2.3}) in the Friedman universe we can confront them with the choice of natural units, including $ \kappa=8\pi G/c^{4}=1$. Such choice formally means that we could not make difference between non-material and material scalar fields. Thus, the equations will give for us the solutions but these solutions should be considered in the subsequent model. That is, from the physical point of view, the solutions will correspond to different representation of gravity.

Let as note also that the cosmological constant $\Lambda$ can be extracted from the constant part of the potential $V(\phi)$, therefore we did not include it into the actions (\ref{2.1}) and (\ref{2.3}).

To describe a homogeneous and isotropic universe we chose the Friedmann-Robertson-Walker (FRW) metric in the form
\begin{equation}\label{2.4}
ds^2=-dt^2+a^2(t)\left(\frac{d r^2}{1-k r^2}+r^2 \left( d\theta^2+\sin^2\theta d\varphi^2\right)\right),
\end{equation}
where $a(t)$ is a scale factor, a constant  $k$ is the indicator of universe's
type:
$ k>0,~k=0,~k<0 $ are associated with closed, spatially flat, open universes, correspondingly.

The cosmological dynamic equations for the GST theory (\ref{2.1}) in a spatially flat $(k=0)$ FRW metric are
\begin{equation}
\label{2.5}
\frac{1}{2}\dot{\varphi}^{2}+V(\varphi)=3FH^{2}+3H\dot{F},
\end{equation}
\begin{equation}
\label{2.6}
\dot{\varphi}^{2}=H\dot{F}-2F\dot{H}-\ddot{F},
\end{equation}
\begin{equation}
\label{2.7}
\ddot{\varphi} + 3H \dot{\varphi} + V'_{\varphi} -3F'_{\varphi}\left(\dot{H}+H^2\right) = 0,
\end{equation}
where a dot represents a derivative with respect to the cosmic time $t$, $H \equiv \dot{a}/a$ denotes the Hubble parameter, $F'_{\varphi} = \partial F/\partial \varphi $ and $\varphi=\int\sqrt{f(\phi)} d\phi$\footnote{We will use the representation over $\varphi$ for the sake of brevity. Nevertheless, the representation in terms of $\phi$ is more convenient, it gives possibility to control crossing of the fantom zone and it's freedom of choice is helpful under performing the integration.},
for $f(\phi)>0$ we have canonical scalar field $\varphi$, for $f(\phi)<0$ we have phantom scalar field $\varphi$ and for $f(\phi)=0$ we have a potential motivated inflation.
The scalar field equation (\ref{2.7}) can be derived from the equations (\ref{2.5})--(\ref{2.6}); for this reason, the equations (\ref{2.5})--(\ref{2.6}) completely describe the cosmological dynamics and we will deal with the GST gravitational equations only
\begin{equation}
\label{2.8}
\frac{f(\phi)}{2}\dot{\phi}^{2}+V(\phi)=3FH^{2}+3H\dot{F},
\end{equation}
\begin{equation}
\label{2.9}
f(\phi)\dot{\phi}^{2}=H\dot{F}-2F\dot{H}-\ddot{F}.
\end{equation}
The gravitational equations (\ref{2.8})-(\ref{2.9}) are represented in terms of $f(\phi), \phi$ notations. We will refer to these equations 
as for {\it GST cosmology} equations.

If $F=1$, equations (\ref{2.8})--(\ref{2.9}) are reduced to those for scalar field Friedmann (inflationary) cosmology
\begin{equation}
\label{2.10}
3H^{2}=\frac{1}{2}\dot{\varphi}^{2}+V(\varphi),
\end{equation}
\begin{equation}
\label{2.11}
\dot{\varphi}^{2}=-2\dot{H}.
\end{equation}
Here, we remember, the scalar field $ \varphi$ is the source of Einstein gravity. We will refer to equations (\ref{2.10})-(\ref{2.11}) as for {\it GR cosmology} equations\footnote{We introduced the terms {\it GST cosmology} and {\it GR cosmology} with the aim to distinguish the gravitational theories background. In both cases we deal with Friedman cosmology because of FRW metric of the spacetime is applied.}.

\section{Conformal connection to GR cosmology}
Let us consider the conformal transformation from GST theory to GR without adopting the scalar field to the canonical form. I.e. we are considering 1D Chiral Cosmological Model (CCM) \cite{Chervon:2014dya}
with the action
\begin{equation}\label{1d-ccm}
 S_{(1D-CCM)} = \int d^4x\sqrt{-g}\left[\frac{R}{2\kappa} -
 \frac{1}{2}h_{11}(\phi)g^{\mu\nu}\partial_{\mu}\phi \partial_{\nu} \phi
 - V(\phi)\right]
\end{equation}
in Einstein frame.

Now we intend to connect the exact solutions of the GST cosmology (\ref{2.1}) to those for GR cosmology \eqref{1d-ccm}. 
To this end we use the conformal transformation\footnote{The quantities in Einstein frame we leave without any indices while the quantities in Jordan frame we marked by ``J" index.}
\begin{equation}
 g_{\mu\nu}(x)=\Omega^2 g^{J}_{\mu\nu}(x),
\end{equation}
with
\begin{equation}
 \Omega=\sqrt{F(\phi)}.
\end{equation}

Using the standard procedure \cite{Fujii:2003pa} one can transform the action \eqref{2.1} to \eqref{1d-ccm}
where
\begin{equation}
 h_{11}(\phi)=\frac{3}{2\kappa}\left(\frac{F'}{F}\right)^2+
 \frac{f(\phi)}{\kappa F},~~ U(\phi)=\frac{V(\phi)}{\kappa F^2}.
\end{equation}
Note, that $h_{11}(\phi)=f(\phi)/\kappa $ in the action (\ref{2.1}) for GR cosmology when $F(\phi)=1$.

Thus we can see, that for transformation from Jordan to Einstein frame we must include Einstein gravitational constant $\kappa $ into the chiral 1D metric and the potential. So the connection between GST and GR cosmologies have the same view as we stated earlier, namely if we set the natural units, including $\kappa=1$.

The metric, conformal to FRW one, became
\begin{equation}
 ds^2=F(\phi) \left[-dt^2+a^2(t)\left(\frac{d r^2}{1-k r^2}+r^2 \left( d\theta^2+\sin^2\theta d\varphi^2\right)\right)\right].
\end{equation}

Thus, knowing the exact form of $F(\phi),~f(\phi),~V(\phi),~H(\phi)$ , we can obtain the solution in GR cosmology.

The Einstein equations for the model $S_{(1D-CCM)}$ can be represented in the form
\begin{equation}
R_{\mu\nu}=h_{11}(\phi)\partial_\mu \phi \partial_\nu \phi +g_{\mu\nu}U(\phi).
\end{equation}

Nonzero components of Ricci tensor are
\begin{equation}
R_{44}=-3\left[ H^2+\dot{H}+\frac{\dot{F}}{2F}H+\frac{\ddot{F}}{2F}-\frac{1}{2}\frac{\dot{F}^2}{F^2}\right],
\end{equation}

 \begin{equation}
  R_{11}=\ddot{a}a+2\dot{a}^2+a^2\frac{\ddot{F}}{2F}+\frac{5}{2}\frac{\dot{F}}{F}\dot{a}a,
\end{equation}

 \begin{equation}
   R_{22}=r^2R_{11},~~R_{33}=r^2 \sin^2 \theta R_{11}.
\end{equation}

Thus, using obtained solution in GST cosmology one can check the corresponding solution in GR cosmology.

\section{GST as the source of deviation from de Sitter stage}\label{De Sitter}

The de Sitter stage with $H=const$ where the source of the accelerated (exponential) expansion is the vacuum energy determined by the cosmological constant $\Lambda$ or a constant scalar field $\phi=const$ is very important model of accelerated universe.

However, within the framework of the inflationary paradigm, a quasi-de Sitter stage with $H\approx const$ is considered, implying quantum fluctuations of the scalar field, leading to the formation of a large-scale structure and gravitational waves in GR cosmology~\cite{Mukhanov:1990me,Liddle} and in STG cosmology~\cite{Faraoni,Fujii:2003pa}. Let us note that the first and correct calculation of quantum generation of GW during a metastable quasi-de Sitter stage (later dubbed inflationary) in GR was made in the work by Starobinsky \cite{Starobinsky:1979ty}, while the first calculation of generation of scalar perturbations in the $R+R^2 $ inflationary model \cite{Starobinsky:1980te} was performed in \cite{Mukhanov:1981xt}.
Scalar perturbations in the first viable inflationary model in GR (the so called "new" inflationary one) were later independently calculated in \cite{Hawking:1982plb,Starobinsky:1982plb,Guth:1982prl}.

Let us consider the GST theory (\ref{2.1}) in FRW spacetime (\ref{2.4}) taking the equations (\ref{2.8})-(\ref{2.9}) as the basic ones. To connect the evolution of the scalar field $\phi$ and  the potential $V(\phi)$ with non-minimal coupling of the field and curvature, which is defined by the function $F(\phi)$, we suggest the $H\&F(\phi)$ ansatz
\begin{equation}
\label{3.12}
H=\lambda {\cal F}[(F(\phi)],
\end{equation}
where $\lambda$ is an arbitrary constant, ${\cal F}[F(\phi)]$ is the function of $F(\phi)$ obeying the property that ${\cal F}[F(\phi)]=1$ when $F=1$. If we wish to make comparison with GR cosmology we chose $F=1$ and we obtain de Sitter stage $H=\lambda$.

Now, we consider the case when ${\cal F}[F(\phi)]=\sqrt{F(\phi)}$. Then the ansatz (\ref{3.12}) is reduced to
\begin{equation}
\label{3.13}
H=\lambda\sqrt{F}, \,\,\,\, \lambda>0.
\end{equation}
Using the ansatz (\ref{3.13}) the equations (\ref{2.8}) -- (\ref{2.9}) can be displayed as
\begin{equation}
\label{3.14}
V(\phi(t))=3\lambda^{2}F^{2}+3\lambda\sqrt{F}\dot{F}+\frac{1}{2}\ddot{F},
\end{equation}
\begin{equation}
\label{3.15}
f(\phi(t))\dot{\phi}^{2}=-\ddot{F}.
\end{equation}

Under the condition $F(\phi)=1$ we turn to GR cosmology and we have the minimal coupling of the (material) scalar field to the curvature. Then, from the equations (\ref{3.13}) -- (\ref{3.15}), we obtain $V=3\lambda^ {2}=\Lambda$, $\phi=const$, $H =\lambda$ and the state parameter $w_ {eff}=-1-2\dot{H}/3H^{2}=- 1$.

Suggesting $F(\phi)\neq 1$, we found the deviation of the scalar field $\phi=\phi (t)$, the potential $V(\phi)$ and the kinetic energy $\dot{\phi}^{2}/2$ from that in GR cosmology at de Sitter stage. Thus, the non-minimal coupling $F(\phi)$ defines the evolution of a scalar field and the deviation of its potential from the flat one.  Also, the non-minimal coupling $F(\phi)$ can drastically change cosmological dynamics relative to GR cosmology (de Sitter expansion in our consideration). As an example of proposed approach we consider the inflation with power-law Hubble parameter.

\section{The inflation with power-law Hubble parameter}

Now, we consider the following coupling function
\begin{equation}
\label{F}
F(t)=\frac{B^{2}}{\lambda^{2}}t^{2n},~~F(t)\equiv F(\phi(t)),
\end{equation}
with corresponding Hubble parameter and scale factor:
\begin{equation}
\label{H}
H(t)=Bt^{n},
\end{equation}
\begin{equation}
\label{a}
a(t)=a_{s}\exp\left(\frac{B}{n+1}t^{n+1}\right),~~~n\neq-1.
\end{equation}

For $-1<n<0$ one has the intermediate inflation~\cite{Barrow:1990vx,Muslimov:1990be} which is faster then power-law inflation $a(t)\propto t^{B}$ ($n=-1$) and slower than de Sitter expansion $a(t)\propto e^{Bt}$ ($n=0$). When $n>0$ we have the other regime of expansion which is faster then de Sitter expansion. In this regime one has $\dot{H}>0$.

Form equations (\ref{3.14})--(\ref{3.15}) we obtain
\begin{equation}
\label{V}
V(\phi(t))=\frac{B^{2}}{\lambda^{2}}\left(3B^{2}t^{4n}+6Bnt^{3n-1}+n(2n-1)t^{2(n-1)}\right),
\end{equation}
\begin{equation}
\label{phi}
f(\phi(t))\dot{\phi}^{2}=-\frac{2n(2n-1)B^{2}}{\lambda^{2}}t^{2(n-1)}.
\end{equation}

Thus, one can generate the exact solutions for any type of the coupling function $F=F(\phi)$ by the choice of the dependence $t=t(\phi)$ from which we also obtain the evolution of a scalar field $\phi=\phi(t)$. For $n=0$ and $B=\lambda$ we return to the de Sitter expansion based on the minimal coupling $F=1$.

Further, we shell consider the exact solutions for some classes of inflationary models with exponential power-law dynamics.

\subsection{The first class of models with $n=1/3$}

For the first class of models with $n=1/3$ from (\ref{V})--(\ref{phi}) we obtain
\begin{equation}
\label{can}
F(t)=\frac{B^{2}}{\lambda^{2}}t^{2/3},~~H(t)=Bt^{1/3},~~a(t)=a_{s}\exp\left(\frac{3B}{4}t^{4/3}\right),
\end{equation}
\begin{equation}
\label{Vcan}
V(t)=\frac{B^{2}}{\lambda^{2}}\left(3B^{2}t^{4/3}-\frac{1}{9}t^{-4/3}+2B\right),
\end{equation}
\begin{equation}
\label{phican}
f(t)\dot{\phi}^{2}=\frac{2B^{2}}{9\lambda^{2}}t^{-4/3}.
\end{equation}
For the following scalar field evolution
\begin{equation}
\phi(t)=\frac{\lambda^{2}}{B^{2}}t^{2/3},
\end{equation}
from equations (\ref{can})--(\ref{phican}) we have
\begin{equation}
V(\phi)=\frac{B^{2}}{\lambda^{2}}\left(\frac{3B^{6}}{\lambda^{4}}\phi^{2}-
\frac{\lambda^{4}}{9B^{4}}\phi^{-2}+2B\right),
\end{equation}
\begin{equation}
F(\phi)=\phi,~~~f(\phi)=\frac{B^{4}}{\lambda^{4}}\phi^{-1},
\end{equation}
that corresponds to the case of Brans-Dicke gravity.

\subsection{The second class of models with $n=1$}

For the second class of models with $n=1$ one has
\begin{equation}
\label{phantom}
F(t)=\frac{B^{2}}{\lambda^{2}}t^{2},~~H(t)=Bt,~~a(t)=a_{s}\exp\left(\frac{B}{2}t^{2}\right),
\end{equation}
\begin{equation}
\label{Vphantom}
V(t)=\frac{B^{2}}{\lambda^{2}}\left(3B^{2}t^{4}+6Bt^{2}+1\right),
\end{equation}
\begin{equation}
\label{phiphantom}
f(t)\dot{\phi}^{2}=-\frac{B^{2}}{\lambda^{2}}.
\end{equation}

For the scalar field $\phi=t$ we have
\begin{equation}
V(\phi)=\frac{B^{2}}{\lambda^{2}}\left(3B^{2}\phi^{4}+6B\phi^{2}+1\right),
\end{equation}
\begin{equation}
F(\phi)=\frac{B^{2}}{\lambda^{2}}\phi^{2},~~~f=-\frac{B^{2}}{\lambda^{2}},
\end{equation}
that corresponds to the case of induced gravity.

For model with non-minimal coupling $F(\phi)=1-\xi\phi^{2}$, where $\xi$ is the dimensionless coupling constant, in the context of second class models, we have
\begin{equation}
V(\phi)=3\lambda^{2}\xi(\xi-2)\phi^{4}-6B\xi\phi^{2}+3\lambda^{2}+6B+\frac{B^{2}}{\lambda^{2}},
\end{equation}
\begin{equation}
\sqrt{\xi}\phi(t)=\pm\left(1-\frac{B^{2}}{\lambda^{2}}t^{2}\right)^{1/2},
\end{equation}
\begin{equation}
f(\phi)=\frac{\xi^{2}\phi^{2}}{\xi\phi^{2}-1}.
\end{equation}

\subsection{The third class of models with $n=1/2$}

In the case of $n=1/2$ we obtain
\begin{equation}
\label{z}
F(t)=\frac{B^{2}}{\lambda^{2}}t,~~H(t)=B\sqrt{t},~~a(t)=a_{s}\exp\left(\frac{2B}{3}t^{3/2}\right),
\end{equation}
\begin{equation}
\label{Vz}
V(t)=\frac{B^{2}}{\lambda^{2}}\left(3B^{2}t^{2}+3B\sqrt{t}\right),
\end{equation}
\begin{equation}
\label{phiz}
f(t)\dot{\phi}^{2}=0.
\end{equation}

For this class of models in the case of induced gravity $F(\phi)=\frac{B^{2}}{\lambda^{2}}\phi^{2}$ or $t=\phi^{2}$ we have the potential
\begin{equation}
V(\phi)=\frac{B^{2}}{\lambda^{2}}\left(3B^{2}\phi^{4}+3B\phi\right).
\end{equation}

For model with non-minimal coupling $F(\phi)=1-\xi\phi^{2}$ we have
\begin{equation}
V(\phi)=3\lambda^{2}\xi^{2}\phi^{4}-6\lambda^{2}\xi\phi^{2}+\frac{3B^{2}}{\lambda}\sqrt{1-\xi\phi^{2}}
+3\lambda^{2}+\frac{B^{2}}{\lambda^{2}},
\end{equation}
\begin{equation}
\sqrt{\xi}\phi(t)=\pm\left(1-\frac{B^{2}}{\lambda^{2}}t\right)^{1/2}.
\end{equation}

Thus, we obtain the new exact solutions for the inflationary models based on the different well known types of a scalar-tensor gravity.

\subsection{The slow-roll regime}

Further, we consider the conditions of slow-roll regime on the basis of the slow-roll parameters
\begin{equation}
\label{flow}
\epsilon=-\frac{\dot{H}}{H^{2}},~~~\delta=\epsilon-\frac{\dot{\epsilon}}{2H\epsilon}=-\frac{\ddot{H}}{2H\dot{H}},~~~~
\xi=\epsilon\delta-\frac{1}{H}\dot{\delta}.
\end{equation}

For $H(t)=Bt^{n}$ with $B<0$ we have the following slow-roll parameters
\begin{equation}
\label{flowEPL}
\epsilon=\frac{n}{B}t^{-(n+1)},~~~\delta=\frac{(n-1)}{2B}t^{-(n+1)},~~~
\xi=-\frac{(n-1)}{2B^{2}}t^{-2(n+1)}.
\end{equation}

Thus, for the values $n=1/3,1,1/2$ the slow-roll parameters are decreasing functions (for $n=1$ one has $\delta=0$, $\xi=0$), therefore, we can always choose the constant $B$ in order to satisfy the conditions $\epsilon\ll1$, $\delta\ll 1$ and $\xi\ll 1$.

The conditions of the slow-roll regime in the case of scalar-tensor gravity
\begin{equation}
|\ddot{\phi}|\ll H|\dot{\phi}|\ll H^{2}|\phi|,
\end{equation}
\begin{equation}
\frac{1}{2}|f(\phi)|\dot{\phi}^{2}\ll |V(\phi)|,
\end{equation}
can be provided when $\frac{\dot{F}}{HF}\ll 1$~\cite{Morris:2001ad}.
For the coupling function (\ref{F}) and the Hubble parameter (\ref{H}) with $B<0$ we have $\delta_{F}=\frac{\dot{F}}{HF}=-2\epsilon$, thus $|\delta_{F}|\ll 1$.

Further, we consider the parameter $\epsilon_{s}$ which also characterizes the possibility of the slow-roll regime in the case of inflation based on the scalar-tensor gravity and can be written as
\cite{DeFelice:2011zh,DeFelice:2011jm}
\begin{equation}
\epsilon_{s}=\epsilon+\frac{1}{2}\delta_{F}+O(\epsilon^{2}).
\end{equation}
For $\delta_{F}=-2\epsilon$ one has $\epsilon_{s}=O(\epsilon^{2})\ll 1$. Thus, it is possible to provide the slow-roll regime for all considered models by the choice of the constant $B$.

\section{Cosmological perturbations}

Further, we shall calculate the parameters of cosmological perturbations to verify the obtained
solutions by the observational constraints.
The method of calculating the parameters of cosmological perturbations in the case of cosmological models based on scalar-tensor gravity can be found, for example, in~\cite{Boisseau:2000pr,Starobinsky:2001xq,DeFelice:2011zh,DeFelice:2011jm}.
In the slow-roll regime we can use the approach outlined in \cite{DeFelice:2011zh,DeFelice:2011jm}.

Firstly, we obtain the functions
\begin{eqnarray}
\label{w1}
&& w_{1}=F,\\
\label{w2}
&& w_{2}=2HF+\dot{F},\\
\label{w3}
&& w_{3}=-9FH^{2}-9H\dot{F}+\frac{3}{2}f(\phi)\dot{\phi}^{2}=
-9FH^{2}-9H\dot{F}+\frac{3}{2}(H\dot{F}-2F\dot{H}-\ddot{F}),\\
\label{w4}
&& w_{4}=F.
\end{eqnarray}
Also, we note that the velocities of the scalar and tensor perturbations for the case of the scalar-tensor gravity are equal to unity ($c_{S}=1$, $c_{T}=1$)~\cite{Pozdeeva:2016cja}.

The power spectrum of the curvature perturbation is given by \cite{DeFelice:2011zh,DeFelice:2011jm}
\begin{equation}
\label{PR}
{\cal P}_{{\rm S}}=\frac{H^{2}}{8\pi^{2}Q_{S}},~~~Q_{S} \equiv  \frac{w_{1}(4w_{1}w_{3}+9w_{2}^{2})}{3w_{2}^{2}}.
\end{equation}

On the crossing of Hubble radius ($k=aH$) the expression $d\ln k$ can be written as
$d\ln k=(H+\frac{\dot{H}}{H})dt=H(1-\epsilon)dt$, where $k$ is the wave number.

In this case, we have the scalar spectral index
\begin{eqnarray}
\label{nR}
n_{{\rm S}}-1\equiv\frac{d\ln{\cal P}_{{\rm S}}}{d\ln k}\bigg|_{k=aH}=\frac{{\cal \dot{P}}_{{\rm S}}}{H(1-\epsilon){\cal P}_{{\rm S}}}\bigg|_{k=aH},
\end{eqnarray}

For tensor perturbations we have the power spectrum \cite{DeFelice:2011zh,DeFelice:2011jm}
\begin{equation}
\label{Pt}
{\cal P}_{{\rm T}}=\frac{H^{2}}{2\pi^{2}Q_{T}},~~~Q_{T}  \equiv  \frac{w_{1}}{4}.
\end{equation}


The spectral index of tensor perturbations is
\begin{eqnarray}
\label{nt}
n_{{\rm T}}\equiv\frac{d\ln{\cal P}_{{\rm T}}}{d\ln k}\bigg|_{k=aH}=\frac{{\cal \dot{P}}_{{\rm T}}}{H(1-\epsilon){\cal P}_{{\rm T}}}\bigg|_{k=aH}.
\end{eqnarray}
The spectral indexes (\ref{nR}) and (\ref{nt}) differs from that given in the works \cite{DeFelice:2011zh,DeFelice:2011jm} differ by a factor of $1/(1-\epsilon)$, since we don't use the slow-roll approximation on this step of calculations. Such a representation of the spectral parameters of cosmological perturbations was used in articles~\cite{Chervon:2004uv,Chervon:2005zz,Chervon:2008zz} to refine its values at the crossing of the Hubble radius for standard inflation with the Einstein gravity.

Tensor-to-scalar ratio is
\begin{equation}
\label{tsr}
r=\frac{{\cal P}_{{\rm T}}}{{\cal P}_{{\rm S}}}=4\frac{Q_{S}}{Q_{T}}.
\end{equation}
Now, we calculate the parameters of cosmological perturbations for the considered models.

\subsection{The parameters of cosmological perturbations for ansatz $H=\lambda\sqrt{F}$}
For the ansatz (\ref{3.12}) we have the velocities and parameters of cosmological perturbations for arbitrary Hubble parameter $H$ in the following form
\begin{equation}
\label{5.13}
{\cal P}_{{\rm S}}=\frac{\lambda^{2}(H^{2}+\dot{H})^{2}}{8\pi^{2}(2\dot{H}^{2}-H\ddot{H})},~~~{\cal P}_{{\rm T}}=\frac{2\lambda^{2}}{\pi^{2}},
\end{equation}
\begin{equation}
\label{5.12}
r=\frac{16(2\dot{H}^{2}-H\ddot{H})}{(H^{2}+\dot{H})^{2}},~~~n_{{\rm T}}=0.
\end{equation}
In terms of the slow-roll parameters we obtain
\begin{equation}
\label{flowPS}
{\cal P}_{{\rm S}}=\frac{\lambda^{2}}{8\pi^{2}\epsilon}\left[\frac{(1-\epsilon)^{2}}{2(\epsilon-\delta)}\right],~~~~~~
r=16\epsilon\left[\frac{2(\epsilon-\delta)}{(1-\epsilon)^{2}}\right],
\end{equation}
\begin{eqnarray}
\label{flownS}
n_{{\rm S}}-1=\frac{\dot{\epsilon}\epsilon\delta-\dot{\delta}\epsilon^{2}-2\dot{\epsilon}\epsilon+\dot{\epsilon}\delta+\epsilon\dot{\delta}}
{(1-\epsilon)^{2}(\epsilon-\delta)\epsilon H}=
\frac{1}{(1-\epsilon)^{2}}\left[2\epsilon\delta-4\epsilon+2\delta+
(1-\epsilon)\left(\frac{\epsilon\delta-\xi}{\epsilon-\delta}\right)\right].
\end{eqnarray}

Thus, for the slow-roll parameters (\ref{flowEPL}) we have
\begin{equation}
\label{5.16}
{\cal P}_{{\rm S}}=\frac{\lambda^{2}(Bt^{n+1}-n)^{2}}{8\pi^{2}n(n+1)},~~~{\cal P}_{{\rm T}}=\frac{2\lambda^{2}}{\pi^{2}}, ~~~r=\frac{16n(n+1)}{(Bt^{n+1}-n)^{2}},
\end{equation}
\begin{equation}
\label{5.17}
n_{{\rm S}}-1=-\frac{2(n+1)Bt^{n+1}}{(Bt^{n+1}-n)^{2}}=-\frac{nr+4\sqrt{n(n+1)r}}{8n},~~~~n_{{\rm T}}=0.
\end{equation}

The e-folds number as function of time is
\begin{equation}
\label{efolds}
N(t)=-\int Hdt=\frac{B}{n+1}t^{n+1}
\end{equation}
and the parameters of cosmological perturbations in terms of the  e-folds number $N$ are
\begin{equation}
\label{rN}
r=\frac{16n(n+1)}{(N(n+1)+n)^{2}},
\end{equation}
\begin{equation}
\label{PN}
{\cal P}_{{\rm S}}=\frac{\lambda^{2}(N(n+1)-n)^{2}}{8\pi^{2}n(n+1)},
\end{equation}
\begin{equation}
\label{nSN}
n_{{\rm S}}-1=-\frac{2(n+1)^{2}N}{(N(n+1)-n)^{2}}.
\end{equation}

Further, we estimate the value of parameter $\lambda$ for three classes of models for $N=60$ and ${\cal P}_{{\rm S}}=2.1\times 10^{-9}$.

I. The first class of models with $n=1/3$

\begin{equation}
r\simeq0.001,~~~{\cal P}_{{\rm T}}\simeq2.4\times 10^{-12},
\end{equation}
\begin{equation}
n_{{\rm S}}-1\simeq-0.033,~~~~\lambda\simeq3.4\times 10^{-6}.
\end{equation}

II. The second class of models $n=1$

\begin{equation}
r\simeq0.002,~~~{\cal P}_{{\rm T}}\simeq4.7\times 10^{-12},
\end{equation}
\begin{equation}
n_{{\rm S}}-1\simeq-0.034,~~~~\lambda\simeq4.8\times 10^{-6}.
\end{equation}

II. The third class of models $n=1/2$

\begin{equation}
r\simeq0.002,~~~{\cal P}_{{\rm T}}\simeq3.1\times 10^{-12},
\end{equation}
\begin{equation}
n_{{\rm S}}-1\simeq-0.034,~~~~\lambda\simeq3.9\times 10^{-6}.
\end{equation}
As one can see, all three classes of models have a good correspondence to
the observational constraints from the PLANCK observations~\cite{Ade:2015xua}
\begin{gather*}
10^{9}{\mathcal{P}}_{S}=2.142\pm0.049, ~~{\mathcal{P}}_{T}=r{\mathcal{P}}_{S}, \\
n_{S}=0.9667\pm0.0040,~~r< 0.112.
\end{gather*}
For Einstein gravity $F=1$ and from (\ref{3.12}) we have the model with cosmological constant
\begin{equation}
\Lambda=3\lambda^{2}\approx10^{-11}
\end{equation}
and with de Sitter dynamics $a(t)\propto\exp(\lambda t)$.

\section{Conclusion and discussions}

In this paper we consider the early cosmology with scalar-tensor gravity as the source of deviation from pure de Sitter expansion, i.e. with nonminimal coupling of a constant scalar field, associated with a primary vacuum, and curvature that leads to its evolution.

For the analysis of cosmological models, we chose the specific connection $H=\lambda\sqrt{F}$ between Hubble parameter and coupling function.

For Hubble parameter $H(t)=Bt^{n}$, which was considered in all models in this article we get the parameters of cosmological perturbations corresponding to observational constraints and estimate the value of cosmological constant on the crossing of Hubble radius.

Also, we note, that the power spectrum of tensor perturbations is constant for any cosmological model based on the scalar-tensor gravity with $H=\lambda\sqrt{F}$.

\section{Acknowledgements}

The authors are grateful to S.Yu. Vernov for the discussion and comments on the material of the article at the
3rd International Winter School-Seminar on Gravity, Astrophysics and Cosmology ``Petrov School", 2017, Kazan.

I.V. Fomin  and S.V. Chervon were supported by RFBR grant 18-52-45016 IND a.
S.V.Chervon is grateful for support by the Program of Competitive Growth of Kazan Federal University.

\end{document}